\documentclass[prd, aps, floatfix, 10pt, twocolumn, superscriptaddress, preprintnumbers]{revtex4-2}

\usepackage[utf8]{inputenc}

\usepackage{xcolor}

\definecolor{lcolor}{rgb}{0.5,0,0}
\definecolor{citcolor}{rgb}{0,0.3,0.0}

\usepackage[breaklinks,colorlinks,urlcolor=blue,citecolor=citcolor,linkcolor=lcolor]{hyperref}

\usepackage{amsmath}
\usepackage{physics}
\usepackage{siunitx}
\usepackage{slashed}

\usepackage{graphicx}

\usepackage{mathtools}

\usepackage{overpic}

\usepackage{subfigure}

\newcommand{\jet}{\text{jet}}

\newcommand{\rt}{\mathbf{r}}
\newcommand{\kt}{\mathbf{k}}
\newcommand{\xt}{\mathbf{x}}

\newcommand{\yt}{\mathbf{y}}

\newcommand{\pt}{\mathbf{p}}
\newcommand{\qt}{\mathbf{q}}

\newcommand{\GeV}{{{\,}\textrm{GeV}}}
\newcommand{\TeV}{{{\,}\textrm{TeV}}}

\newcommand*\diff{\mathop{}\!\mathrm{d}}

\graphicspath{ {figures/} }

\begin{document}

\title{Transverse Energy--Energy Correlators at Small $x$ for Photon--Hadron Production}

\author{Zhong-Bo Kang}
\email{zkang@physics.ucla.edu}
\affiliation{Department of Physics and Astronomy, University of California, Los Angeles, CA 90095, USA}
\affiliation{Mani L. Bhaumik Institute for Theoretical Physics, University of California, Los Angeles, CA 90095, USA}
\affiliation{Center for Frontiers in Nuclear Science, Stony Brook University, Stony Brook, NY 11794, USA}

\author{Robert Kao}
\email{rqk@ucla.edu}
\affiliation{Department of Physics and Astronomy, University of California, Los Angeles, CA 90095, USA}
\affiliation{Mani L. Bhaumik Institute for Theoretical Physics, University of California, Los Angeles, CA 90095, USA}

\author{Meijian Li}
\email{meijian.li@usc.es}
\affiliation{Instituto Galego de Fisica de Altas Enerxias (IGFAE), Universidade de Santiago de Compostela, E-15782 Galicia, Spain}

\author{Jani Penttala}
\email{janipenttala@physics.ucla.edu}
\affiliation{Department of Physics and Astronomy, University of California, Los Angeles, CA 90095, USA}
\affiliation{Mani L. Bhaumik Institute for Theoretical Physics, University of California, Los Angeles, CA 90095, USA}

\begin{abstract}

We study the transverse energy--energy correlator (TEEC) observable in photon--hadron and photon--jet production in p+p and p+A collisions at small $x$.
We derive the relevant expressions in the high-energy limit of the scattering where the dipole picture is applicable and show how the dependence on the fragmentation function of the hadron cancels due to the momentum-sum rule.
The nonperturbative scattering with the target nucleus is expressed in terms of the dipole amplitude, which also describes nonlinear gluon saturation effects.
The TEEC observable is computed in the RHIC and LHC kinematics, and we show that it can be sensitive to the dipole amplitude, making it a potentially good observable for studying saturation effects.

\end{abstract}

\maketitle

\section{Introduction}

The gluon density of nucleons has been observed to grow rapidly with increasing energy~\cite{ZEUS:2002xjx}.
In the high-energy limit of QCD, nonlinear gluon recombination effects start to become important, taming down the growth of the gluon density.
This gluon saturation phenomenon can be understood using the color-glass condensate effective field theory~\cite{Gelis:2010nm,Weigert:2005us} where the nonlinear evolution of the nuclear gluon field is given by the Jalilian-Marian--Iancu--McLerran--Weigert--Leonidov--Kovner (JIMWLK)~\cite{Iancu:2000hn,Iancu:2001ad,Iancu:2001md,Ferreiro:2001qy,Jalilian-Marian:1996mkd,Jalilian-Marian:1997jhx,Jalilian-Marian:1997qno} equation.
Because of the numerical complexity in solving the JIMWLK equation, the simpler Balitsky--Kovchegov equation~\cite{Balitsky:1995ub,Kovchegov:1999yj} is often used instead.

While gluon saturation is well established from the theoretical point of view, it has been challenging to find clear signs of saturation in the experimental data.
For saturation effects to be important, the experimental observable has to be sensitive to momentum scales of the order of the saturation scale $Q_s \sim \mathcal{O}(\SI{1}{GeV})$.
This makes it challenging to distinguish gluon saturation from other non-perturbative effects that might be present at such low momentum scales.
Thus, choosing a suitable observable to study saturation is of utmost importance.

In this paper, we consider the transverse energy--energy correlator (TEEC) observable as a potential probe for saturation effects.
Energy--energy correlators (EECs) are infrared-safe event-shape observables that were originally designed for electron--positron colliders~\cite{Basham:1978bw,Basham:1978zq}, and TEEC was later developed as a more suitable observable for colliders involving initial hadrons~\cite{Ali:1984yp}.
The main advantage of EEC and TEEC observables is their insensitivity to non-perturbative hadronization effects in the final state, allowing one to have a sharper focus on the other parts of the collision process.

TEEC has gained a lot of interest in  the recent years~\cite{ATLAS:2015yaa, ATLAS:2017qir,CMS:2024mlf,Gao:2019ojf,Gao:2023ivm,Kang:2024otf} and has also been considered at small $x$ for electron--hadron production in deep inelastic scattering~\cite{Kang:2023oqj}. 
In addition, one can also consider the asymmetry of the TEEC (ATEEC)~\cite{Ali:1984yp,Ali:2012rn} as asymmetries in EECs for $e^+e^-$ have previously been found to be advantageous in certain ways, such as eliminating linear errors in a phase-space cutoff parameter~\cite{Ali:1984gzn}, exhibiting much smaller $\mathcal{O}(\alpha_{s}^2)$ corrections~\cite{Ali:1984yp}, and sensitivity to gluon emission in 3-jet events~\cite{JADE:1984taa}, and asymmetries in TEECs may have analogous desirable qualities in hadron collisions.

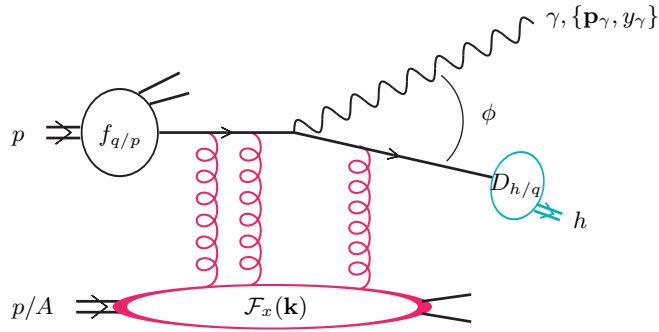
\begin{figure*}[thp!]
    \centering
    \begin{overpic}[width=0.4\textwidth]{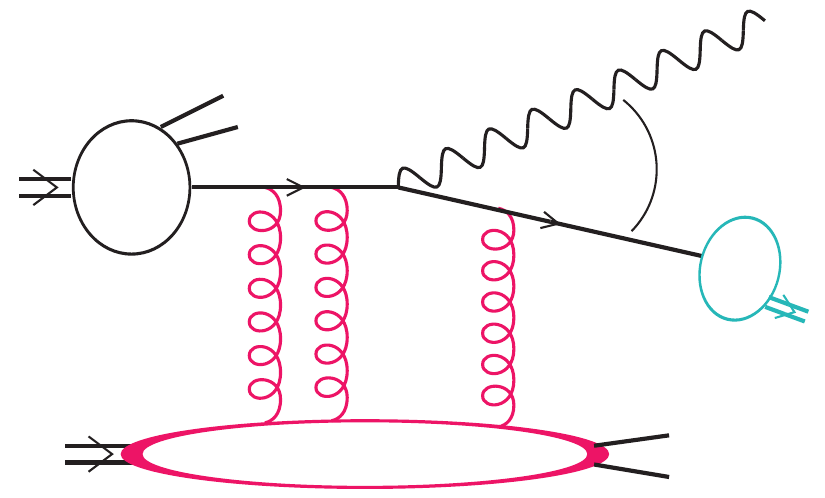}
     \put (-4,36) {$p$}
     \put (12,36) {$f_{q/p}$}
     \put (95,58) {$\gamma, \{\pt_\gamma,  y_\gamma\}$}
    \put (83,40) {$\phi$}
     \put (-4,4) {$p/A$}
     \put (39,4) {$\mathcal F_x(\kt)$}
     \put (84.5,27) {$D_{h/q}$}
     \put (100,20) {$h$}
    \end{overpic}
    \caption{An illustration of the $p+p/A\to \gamma+h+X$ processes in the p+p (or p+A) collisions in the small $x$ regime.
    }
    \label{fig:DIS}
\end{figure*}

We consider the TEEC observable for photon--hadron and photon--jet production in p+p and p+A collisions in the small-$x$ regime. As illustrated in Fig.~\ref{fig:DIS}, at leading order, an incoming parton from the proton scatters off the gluon shock wave (representing multiple gluon exchanges), producing a photon and a final-state parton which subsequently fragments into a hadron or a jet.
There are two main advantages for considering processes with a photon in the final state.
First, measuring the photon momentum allows one to focus on a specific phase space of the collision where the saturation effects are expected to be more pronounced.
Specifically, considering photons produced in the forward direction ensures that the energy of the parton--target collision is high enough for the dipole picture to be applicable, and limiting the photon transverse momentum to be not too high makes the process more sensitive to saturation~\cite{Gelis:2002ki,Jalilian-Marian:2005qbq,Jalilian-Marian:2005tod,Jalilian-Marian:2012wwi,Kopeliovich:2007fv,Kopeliovich:2007sd,Kopeliovich:2007yva,Kopeliovich:2009yw,Benic:2017znu,Benic:2022ixp,Kolbe:2020tlq,Benic:2018hvb}.
Second, photon--hadron production is computationally simpler than hadron--hadron production, for example, which involves higher-order Wilson line correlators~\cite{Dominguez:2011wm,Lappi:2012nh}. 
This means that we can compute the process away from the back-to-back limit that is often employed to make the computation numerically easier but also requires a resummation of large Sudakov logarithms~\cite{Gao:2023ivm,Kang:2024otf}.
Thus, analyzing the numerical results in our case is simpler as
we do not need to introduce additional nonperturbative components related to the Sudakov logarithms. We note a parallel study on photon--hadron TEEC from a Monte Carlo perspective in Ref.~\cite{Ganguli:2025aqa}. 

The paper is organized as follows.
In Sec.~\ref{sec:theory}, we derive the relevant expression for photon--hadron and photon--jet TEEC in the dipole picture.
The numerical results using these expressions are shown in Sec.~\ref{sec:numerics} where we also discuss the role of TEEC as a probe for gluon saturation.
Finally, the results are summarized in Sec.~\ref{sec:conclusions}.

\section{Theoretical formalism}
\label{sec:theory}

In this section, we derive the relevant expression for the TEEC observable in photon--hadron and photon--jet production in p+p and p+A collisions at small~$x$. We also present the details of the dipole amplitude employed in our numerical calculations.

\subsection{Transverse energy--energy correlator}
The TEEC for photon+hadron ($\gamma+h$) production in p+A collisions can be written as
\begin{multline}
\label{eq:TEEC}
    \frac{\dd{\Sigma^{\gamma +h }}}{\dd{y_\gamma} \dd[2]{\pt_\gamma} \dd{\tau}}
    = \sum_h \int \dd{y_h} \dd[2]{\pt_h} 
      \frac{\dd{\sigma^{p+A \to \gamma + h +X}}}{\dd{y_\gamma} \dd[2]{\pt_\gamma}\dd{y_h} \dd[2]{\pt_h}}\\
    \times \frac{E_{\gamma T} E_{hT}}{E_{\gamma T} \qty( \sum_{h'} E_{h' T} )}
    \delta \qty(\tau - \frac{1 + \cos \phi}{2} )\;,
\end{multline}
where $\pt_l$ and $y_l$ represent the transverse momentum and the rapidity of the particle $l$, respectively. The transverse energy is defined as $E_{lT} = \sqrt{\pt_l^2 + m_l^2}$.
The azimuthal angle between the photon and the hadron is denoted by
$\phi=\angle (\pt_\gamma,\pt_h )$, and $\sum_h$ represents a summation over all outgoing hadrons.  Additionally, we use $p_T^l = |\pt_l|$ interchangeably throughout this paper.
The variable $\tau$ is related to the azimuthal angle via
\begin{equation}
    \tau = \frac{1 + \cos \phi}{2} \;.
\end{equation}
In this parametrization, the back-to-back region ($\phi \to \pi$) corresponds to $\tau \to 0$, while the collinear limit ($\phi \to 0$) corresponds to $\tau \to 1$.
We will assume that the transverse momentum is large enough for the particle masses to be negligible and write $E_{lT} = \abs{\pt_l}$.

The cross section for $p+A \to \gamma + h + X$ can be written in terms of partonic cross sections for $a + A \to a + \gamma + X$ as
\begin{multline}\label{eq:cross_pA_photonh}
    \frac{\dd{\sigma^{p+A \to \gamma + h +X}}}{\dd{y_\gamma} \dd[2]{\pt_\gamma}\dd{y_h} \dd[2]{\pt_h}}
    =
      \sum_a \int_{z_{\min}}^1 \frac{\dd{z}}{z^2} D_{h/a}(z, \mu^2)\\  
   \times \int_{x_{\min}}^1 \dd{x_p} f_{a/p}(x_p, \mu^2)
    \frac{\dd{\sigma^{a+A \to a + \gamma + X}}}{\dd{y_\gamma} \dd[2]{\pt_\gamma}\dd{y_a} \dd[2]{\pt_a}}\;,
\end{multline}
where $a$ is a parton from the proton, $x_p$ the momentum fraction of the initial parton $a$ relative to the proton with a lower bound $x_{\min}=p^+_\gamma/P_p^+ $, and we are working in a frame where the proton has a large plus momentum $P_p^+$. 
The parton distribution function (PDF) of the proton is written as $f_{a/p}$, with $\mu^2$ being the typical momentum scale of the process. 
The parton-to-hadron fragmentation function is written as $ D_{h/a}$, and
here we are working in the collinear factorization approach for fragmentation such that the momenta of the hadron and the final parton are related by
$p_h = z \, p_a$. 
The momentum fraction $z$ has a lower bound $z_{\min}=p^+_h/(P_p^+ - p_\gamma^+)$ determined by the kinematics of the process~\cite{Jalilian-Marian:2005qbq}.
We have neglected the hadron mass so that the rapidity of the outgoing parton and the hadron are the same, $y_a=y_h$. Note that, throughout this paper, we work at the leading order in the dipole picture. As a result, the initial parton $a$ in the PDF $f_{a/p}$ is the same as the parton flavor in the fragmentation function $D_{h/a}$, i.e., a quark or an antiquark.

The TEEC kernel in Eq.~\eqref{eq:TEEC} can be expressed as $E_{\gamma T} E_{hT}/[E_{\gamma T} \qty( \sum_{h'} E_{h' T} )]= |\pt_h|/|\pt_a|=z $.
In particular, when inserting Eq.~\eqref{eq:cross_pA_photonh} into Eq.~\eqref{eq:TEEC} and writing the integration of the hadron momentum in terms of the partonic momentum, the lower bound of $z$ becomes $0$. The $z$ integration becomes the momentum-sum rule
\begin{equation}
  \sum_h  \int_0^1 \dd{z} z D_{h/a}(z, \mu^2) = 1\;,
\end{equation}
and
therefore the dependence on the fragmentation function vanishes.
We can then simplify the expression in Eq.~\eqref{eq:TEEC} to
\begin{equation}
\label{eq:TEEC_simplified}
\begin{split}
    &\frac{\dd{\Sigma^{\gamma +h }}}{\dd{y_\gamma} \dd{\pt_\gamma^2} \dd{\tau}}\\
    =& \pi \sum_{a} \int \dd{y_a} \dd[2]{\pt_a} \int_{x_{\min}}^1 \dd{x_p}
    f_{a/p}(x_p, \mu^2)\\
  &\times   \frac{\dd{\sigma^{a+A \to a+\gamma+X }}}{\dd{y_\gamma} \dd[2]{\pt_\gamma}\dd{y_a} \dd[2]{\pt_a}}
    \delta \qty(\tau - \frac{1 + \cos \phi}{2} )\\
    =&  \frac{2\pi}{\sqrt{\tau(1-\tau)}} \sum_{a} \int \dd{y_a}\int \dd{\abs{\pt_a}}\abs{\pt_a}
     \int_{x_{\min}}^1 \dd{x_p}
    \\&
    \times 
    f_{a/p}(x_p, \mu^2)
    \frac{\dd{\sigma^{a+A \to a+\gamma + X}}}{\dd{y_\gamma} \dd[2]{\pt_\gamma}\dd{y_a} \dd[2]{\pt_a}} \Bigg|_{\tau = (1+\cos \phi)/2}.
\end{split}
\end{equation}
Note that as we are using the collinear approximation for hadronization, the angle $\phi$ is also the azimuthal angle between the photon and the parton, i.e., $\phi=\angle (\pt_\gamma,\pt_a )$.
For quarks (and likewise for antiquarks), we can write the partonic cross section as~\cite{Jalilian-Marian:2005qbq,Dominguez:2011wm}
\begin{multline}
\label{eq:quark_cross_section}
    \frac{\dd{\sigma^{q+A \to q+ \gamma+ X}}}{\dd{y_q}\dd{y_\gamma} \dd[2]{\pt_q} \dd[2]{\pt_\gamma}}
    = \delta(1 - z_q - z_\gamma) \frac{2e_q^2 \alpha_\text{em}}{(2\pi)^2} 
    \\
    \times z_q  \qty[1+z_q^2]
    \frac{(\pt_q + \pt_\gamma)^2 z_\gamma^2}{\pt_\gamma^2 \qty[ z_\gamma \pt_q  - z_q \pt_\gamma ]^2}
    S_\perp
     \mathcal{F}_{x_A}(\pt_q + \pt_\gamma) \;,
\end{multline}
where we have denoted by $z_\gamma = p_\gamma^+ / (x_p P_p^+)$ and $z_q = p_q^+  / (x_p P_p^+)$ the momentum fractions of the outgoing particles with respect to the initial quark, $e_q$ is the fractional charge of the quark, and $S_\perp$ is the transverse size of the target.
The (Fourier-transformed) dipole amplitude $\mathcal{F}$ is given by
\begin{equation}
\label{eq:F}
    \mathcal{F}_x(\qt) =  \int \frac{ \dd[2]{\rt}}{(2\pi)^2} 
    e^{-i \qt \vdot \rt}
    S_x(\xt, \yt),
\end{equation}
where $S_x(\xt, \yt)$ is the $S$-matrix for dipole--target scattering with transverse coordinates $\xt$ and $\yt$, with $\rt = \xt - \yt$ being the dipole size.

Substituting Eq.~\eqref{eq:quark_cross_section} into Eq.~\eqref{eq:TEEC_simplified}, the integral over $x_p$ is evaluated using the delta function from the former.
This results in 
\begin{equation}
\label{eq:TEEC_hadron}
\begin{split}
  & \text{TEEC}^{\gamma + h}(\tau) \equiv
    \frac{\dd{\Sigma^{\gamma +h }}}{\dd{y_\gamma} \dd{\pt_\gamma^2} \dd{\tau}}\\
   & =  \frac{2\pi}{\sqrt{\tau(1-\tau)}} \sum_{a} \int \dd{y_a}\int \dd{\abs{\pt_a}}\abs{\pt_a}
     \\&
    \times  x_p
    f_{a/p}(x_p, \mu^2) \frac{2e_a^2 \alpha_\text{em}}{(2\pi)^2} \\&
    \times z_a  \qty[1+z_a^2]
    \frac{(\pt_a + \pt_\gamma)^2 z_\gamma^2}{\pt_\gamma^2 \qty[ z_\gamma \pt_a  - z_a \pt_\gamma ]^2}
    S_\perp
     \mathcal{F}_{x_A}(\pt_a + \pt_\gamma)\;,
\end{split}
\end{equation}
where the variables appearing in the integrand can be written as
\begin{align}
\begin{split}
    x_p &= \frac{\abs{\pt_a}e^{y_a} + \abs{\pt_\gamma}e^{y_\gamma}}{\sqrt{s}}\,,
    \quad
    \cos \phi = 2\tau -1\,, \\
    z_a &= \frac{\abs{\pt_a}e^{y_a}}{\abs{\pt_a}e^{y_a} + \abs{\pt_\gamma}e^{y_\gamma}}\,,
     \quad
    z_\gamma =  \frac{ \abs{\pt_\gamma}e^{y_\gamma}}{\abs{\pt_a}e^{y_a} + \abs{\pt_\gamma}e^{y_\gamma}}\;,
    \end{split}
\end{align}
and in the lab frame
$P_p^+ = P_A^- = \sqrt{s/2}$.
For the factorization scale $\mu$ and the small-$x$ scale used in the dipole amplitude $\mathcal{F}_x$, we choose:
\begin{align}
    \mu^2 = \pt_\gamma^2\,,
    \qquad
    x_A = \frac{\abs{\pt_a}e^{-y_a} + \abs{\pt_\gamma}e^{-y_\gamma}}{\sqrt{s}}\;.
\end{align}

In addition to the photon+hadron TEEC we can also consider the photon+jet ($\gamma + \jet$) TEEC.
In this case, hadronization is not relevant, and the TEEC is expressed differentially in terms of the jet momentum.
At leading order, jets are identified with the outgoing parton $a$, and the corresponding TEEC can be read from Eq.~\eqref{eq:TEEC_hadron} as
\begin{equation}
\label{eq:TEEC_jet}
\begin{split}
  & \text{TEEC}^{\gamma + \jet} (\tau)\equiv
   \frac{\dd{\Sigma^{\gamma + \jet}}}{\dd{y_\gamma} \dd{\pt_\gamma^2} \dd{y_\jet} \dd{\pt_\jet^2} \dd{\tau}}\\
    &=  \frac{\pi}{\sqrt{\tau(1-\tau)}} 
    \sum_{a}   x_p
    f_{a/p}(x_p, \mu^2) \frac{2e_a^2 \alpha_\text{em}}{(2\pi)^2} \\&
    \times z_a  \qty[1+z_a^2]
    \frac{(\pt_a + \pt_\gamma)^2 z_\gamma^2}{\pt_\gamma^2 \qty[ z_\gamma \pt_a  - z_a \pt_\gamma ]^2}
    S_\perp
     \mathcal{F}_{x_A}(\pt_a + \pt_\gamma)\;.
\end{split}
\end{equation}
At leading order, where the jet consists of a single parton, we have $\pt_a = \pt_\jet$ and  $y_a = y_\jet$. Note that here we do not integrate over the momentum of the jet, in contrast to the $\gamma + h$ case.
Equation~\eqref{eq:TEEC_jet} is essentially the same as the differential cross section for $\gamma + \jet$ production as the transverse momentum weight appearing in TEEC is identically one, but at higher orders the expressions would be different.

To study the differences between protons and heavy nuclei, we also define the nuclear modification factor for both the $\gamma+h$ and the $\gamma+\jet$ TEECs:
\begin{equation}\label{eq:RpA_def}
    R_{pA}(\tau) = \frac{1}{A} 
     \frac{\text{TEEC}_A (\tau)}
{\text{TEEC}_p (\tau)}\;.
\end{equation}
The subscript of TEEC, $A$ or $p$, denotes whether the target is a heavy ion ($A$) or a proton ($p$), respectively.
The normalization factor $1/A$ has been chosen such that without saturation effects this ratio should be close to 1.

In addition, we compute the asymmetry in the TEEC between $\phi$ and $\pi - \phi$~\cite{Ali:1984yp,Ali:2012rn}, known as the ATEEC, which is defined as
\begin{align}\label{eq:ATEEC}
    \text{ATEEC}(\tau)\equiv
    \frac{\diff \Sigma^{\text{asym}} }{\diff\tau}
    \equiv
    \left.
    \frac{\diff \Sigma }{\diff\tau'}
    \right\vert_{\tau'=\tau}
    -\left.
    \frac{\diff \Sigma }{\diff\tau'}\right\vert_{\tau'=1-\tau}
    \;.
\end{align}

\subsection{Model inputs}

The $S$-matrix $S_x(\xt,\yt)$ in Eq.~\eqref{eq:F} can be modeled using the color-glass condensate effective field theory~\cite{McLerran:1993ka, McLerran:1993ni, McLerran:1994vd}.
We will assume that it depends only on the dipole size and write
$S_x(\abs{\xt -\yt}) \equiv S_x(\xt, \yt)$.
Consequently, the function $\mathcal{F}$ in Eq.~\eqref{eq:F} also depends only on the magnitude of the momentum: $\mathcal F(\qt) =\mathcal F(|\qt|) $. 
The dependence of $S_x$ on the $x$-variable is given by the running-coupling Balitsky--Kovchegov (rcBK) equation~\cite{Balitsky:1995ub,Kovchegov:1999yj}
\begin{multline}
    \frac{\partial S_x(\abs{\rt})}{\partial \log 1/x} = 
    \int d^{2}{\rt'} \, K^{\rm run}({\rt}, {\rt'})\\
    \times \qty[S_x(|{\rt'}|)  S_x(|{\rt - \rt'}|) - S_x(\abs{\rt})]\;.
\end{multline}
The running-coupling kernel in the Balitsky prescription~\cite{Balitsky:2006wa} reads
\begin{align}
\begin{split}
    K^{\rm run}({\rt}, {\rt'})=&\frac{N_{c}  \alpha_{s}(\rt^2)}{2\pi^{2}} 
    \times\left[\frac{\rt^2}{ \rt^{\prime 2} (\rt -\rt')^{2}} \right.\\
   &+ \frac{1}{\rt^{\prime 2}}\qty(\frac{\alpha_{s}(\rt^{\prime 2})}{\alpha_{s}((\rt -\rt')^{2})}-1) \\
   &\left.+ \frac{1}{(\rt -\rt')^{2}}\qty(\frac{\alpha_{s}((\rt -\rt')^{2})}{\alpha_{s}(\rt^{\prime 2})}-1) \right]\;,
   \end{split}
\end{align}
with the coordinate-space running coupling
\begin{equation}
     \alpha_s(r^2) = \frac{12 \pi}{(33-2 N_f) \log( \frac{4 C^2}{r^2 \Lambda_\text{QCD}^2} )}\;.
\end{equation}
Here, $N_f=3$, $\Lambda_\text{QCD} = \SI{0.241}{GeV}$, and $C^2$ is a free parameter that controls the dependence of the coupling on the length scale $r$.
For the initial condition of the rcBK equation we use the model~\cite{Casuga:2023dcf}
\begin{multline}\label{eq:MVe}
    S_{x_0}(r) =\\ \exp\qty[-\frac{\qty(r^2  Q^2_{s0})^{\gamma}}{4}\log(\frac{1}{\Lambda_{\rm QCD}  r} + e \cdot e_c )]\;.
\end{multline}
For the proton target, we take the median values of the parameters $C^2$, $Q_{s0}^2$, $e_c$, and $S_\perp(=\sigma_0 / 2)$ from the 4-parameter fit (with $\gamma=1$ fixed) in~\cite{Casuga:2023dcf}.
For nuclear targets, we modify the saturation scale and the transverse size by~\cite{Kang:2023oqj}
\begin{align}\label{eq:Qs_A}
    Q_{s0,A}^2 = c A^{1/3} Q_{s0}^2\;,
    \qquad
    S_{\perp, A} = A^{2/3} S_\perp / c\;.
\end{align}
The parameter $c$ takes into account the uncertainty in the nuclear geometry, and we vary it between $0.5 < c < 1.0$~\cite{Dusling:2009ni,Tong:2022zwp}. We do not include a corresponding uncertainty for the proton case, as the model inputs are fixed. The geometric uncertainty will be reflected in the nuclear modification factor.

For the proton PDF, we use the CT18NLO~\cite{Hou:2019qau} set via the LHAPDF~\cite{Buckley:2014ana} library.

\section{Numerical results}
\label{sec:numerics}

We consider two representative kinematic regimes that correspond to the typical phase space probed by the RHIC~\cite{Okawa:2023asr, PHENIX:2018trr} and LHC\cite{ALICE:2020atx, LHCb:2021vww} experiments: 
    \begin{itemize}
        \item RHIC: $p^\gamma_T=5 \GeV $, $y_\gamma=2$, $p^\jet_T=5 \GeV $, $y_\jet=2$, and $\sqrt{s} =200 \GeV$;
        \item LHC: $p^\gamma_T=10 \GeV $, $y_\gamma=4$, $p^\jet_T=10 \GeV $, $y_\jet=4$, and $\sqrt{s} =5.02 \TeV$.
    \end{itemize}

In the numerical evaluation of the photon+hadron TEEC, we perform a numerical integration using adaptive cubature method
 following Eq.~\eqref{eq:TEEC_hadron}. The integration ranges of $y_a$ and $\abs{\pt_a}$ are constrained by momentum conservation, which requires all relevant momentum fractions—$z_\gamma$, $z_q$, and $x_p$—to remain within the physical interval $(0,1)$.

As in the usual small-$x$ implementation, the dipole amplitude $\mathcal F_x(\qt)$ is obtained by carrying out the Fourier transform [as in Eq.~\eqref{eq:F}] of the numerically evolved dipole $S$-matrix $S_x(\rt)$ according to the rcBK equation. 
However, this procedure can be challenging due to the rapid oscillatory nature of the Bessel function, and unphysical oscillations in $\abs{\qt}$ at high momentum often emerge~\cite{Lappi:2013zma, Giraud:2016lgg, Shi:2021hwx, Casuga:2023dcf}. 
To address this issue, we first fit the evolved $S_x(\rt)$ at each $x$ using Eq.~\eqref{eq:MVe}, treating $Q_{s0}$, $\Lambda_{\text{QCD}}$, $\gamma$, and $ e_c $ as free parameters. 
 The Fourier transform is then applied to the fitted $S_x(\rt)$, yielding a smooth and positive $\mathcal{F}_x(\qt)$.
An alternative method, proposed in Ref.~\cite{Shi:2021hwx}, involves directly fitting $\mathcal{F}_x(\qt)$ to a power law at large $\abs{\qt}$ to suppress oscillations.

We examined scale uncertainties for the p+p case, following standard practice in perturbative QCD computations. At LO, the only scale dependence arises from the factorization scale $\mu$ in the proton PDF, with the default choice $\mu = p_T^\gamma$. To estimate the uncertainty, we varied this scale to $\mu = 0.5 p_T^\gamma$ and $\mu = 2 p_T^\gamma$, and found that the TEEC changes by about 4\%. Beyond LO, real gluon emission introduces explicit dependence on the strong coupling $\alpha_s(\mu)$, and the associated scale variation would naturally lead to larger uncertainties, especially in the relatively small transverse-momentum region considered here. A detailed assessment of scale uncertainties in the TEEC for p+p collisions will be left to future work beyond LO.

\begin{figure*}[htp!]
  \centering 
  \subfigure[
    RHIC
    \label{fig:RHIC_jet_TEEC}
  ]{ \includegraphics[width=0.4377\textwidth]{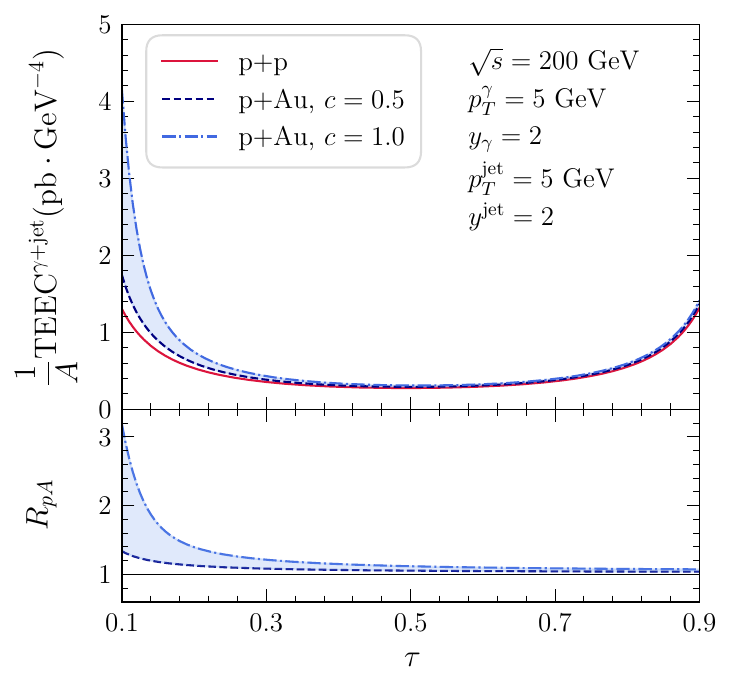} }\qquad
  \subfigure[
    LHC
    \label{fig:LHC_jet_TEEC}
  ]{ \includegraphics[width=0.45\textwidth]
    {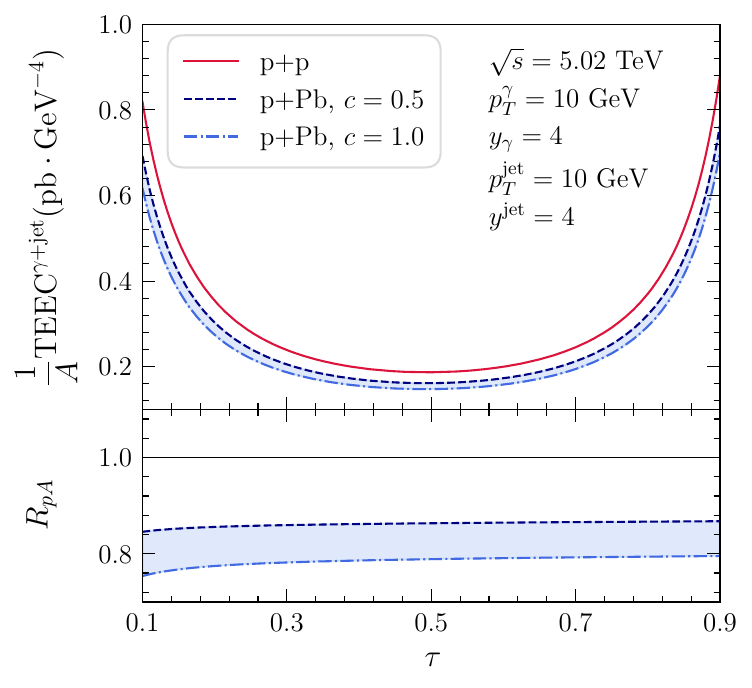} }
  \caption{The photon--jet TEEC, as defined in Eq.~\eqref{eq:TEEC_jet}, and their ratios, as defined in Eq.~\eqref{eq:RpA_def}, as a function of $\tau$.
  }
  \label{fig:TEEC_photon_jet}
\end{figure*}

\subsection{Photon + jet}
\label{sec:numerics_jet}

We calculate the $\gamma+\text{jet}$ TEEC, as defined in Eq.~\eqref{eq:TEEC_jet}, along with the associated nuclear modification ratio $R_{pA}$ in the selected RHIC and LHC kinematic regions. The results are shown in Fig.~\ref{fig:TEEC_photon_jet}.
One can see that the TEEC curves, for both p+p, and p+Au(Pb), exhibit a similar overall pattern, increasing as they approach $\tau=0$ and $\tau=1$, and reaching a minimum around $\tau=0.5$, as indicated by the $1/\sqrt{\tau(1-\tau)}$ factor in Eq.~\eqref{eq:TEEC_jet}.
Here, we focus on the regime $\tau=0.1 \sim 0.9$ where our calculation  applies. The regions near $\tau = 0$ and $\tau = 1$ correspond, respectively, to the back-to-back~\cite{Gao:2023ivm} and collinear~\cite{Dixon:2019uzg} limits of $\gamma$+jet production, and require additional resummation of large logarithmic contributions in the back-to-back limit, this can be achieved using TMD resummation~\cite{Gao:2019ojf,Gao:2023ivm,Kang:2023oqj,Kang:2024otf}, while in the collinear limit one has to resum collinear emissions of final-state particles~\cite{Dixon:2019uzg,Barata:2024wsu}.
While we note that these external regions, especially the back-to-back limit, can also be interesting for studying the nuclear structure, we leave the combination of the resummation techniques with the dipole picture for future work.

In the nuclear modification factor $R_{pA}$, the effect from the $\tau$-dependent overall factor no longer contributes, and the dominant $\tau$ dependence enters through the dipole amplitude $\mathcal F_{x_A}(k)$ where
\begin{equation}
    k=\qty[|\pt_\gamma|^2 +|\pt_\jet|^2 +2(2\tau-1)|\pt_\gamma||\pt_\jet| ]^{1/2}\;.
\end{equation}
Consequently, larger $\tau$ corresponds to a larger transverse momentum transfer in the dipole--target cross section. 
We find that the $R_{pA}$ shows different behaviors in Fig.~\ref{fig:RHIC_jet_TEEC} at the RHIC energy scale and in Fig.~\ref{fig:LHC_jet_TEEC} at the LHC energy scale. In the former case, the $R_{pA}$ is above 1 throughout the probed $\tau$ regime, and decreasing at larger $\tau$, whereas in the latter case, the $R_{pA}$ is below 1 and slightly increases towards larger $\tau$.

\begin{figure*}[htp!]
  \centering
\includegraphics[width=0.44\textwidth]{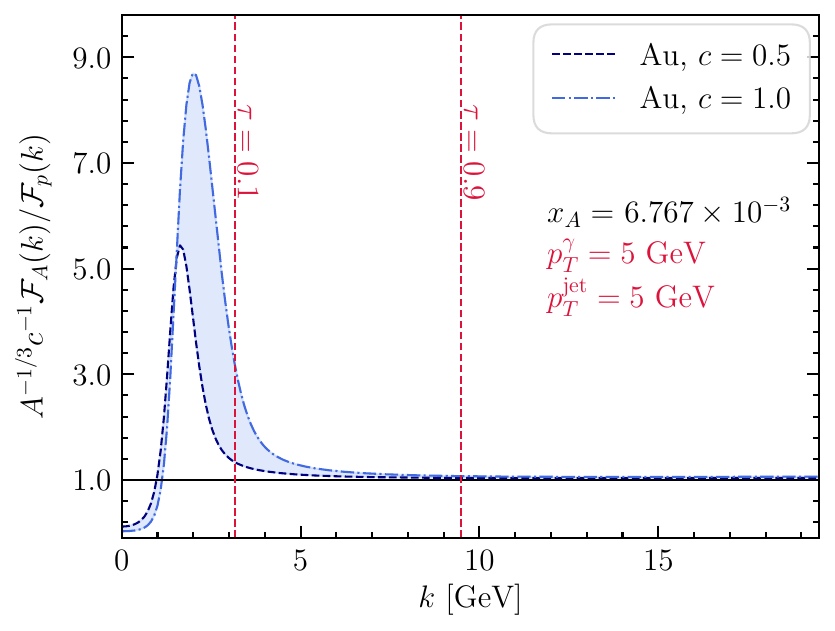}
\qquad
\includegraphics[width=0.44\textwidth]{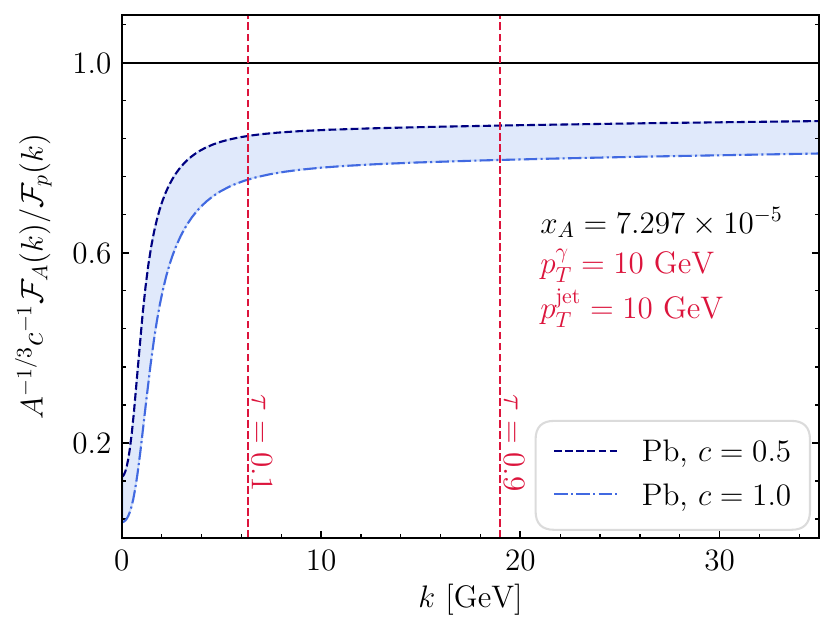}
\caption{The ratio of nucleus--proton dipole amplitude as defined in Eq.~\eqref{eq:R_Fkx}.}
  \label{fig:R_Fkx}
\end{figure*}

To further understand the underlying mechanism of the obtained $R_{pA}$, we examine its relation to the dipole amplitude and in a broader $p_T$ regime.
We note that in calculating the $\gamma+\jet$ TEEC, the nuclear modification factor defined in Eq.~\eqref{eq:RpA_def} reduces to a weighted ratio of the dipole amplitudes between the nucleus and the proton targets,
\begin{equation}\label{eq:R_Fkx}
    R_{pA}^{\gamma+\jet} (\tau) = \frac{1}{A} 
   \frac{S_{\perp,A}}{S_{\perp}}
   \frac{\mathcal F_A(k)}{\mathcal F_p(k)}
   = 
   \frac{1}{c A^{1/3}}
   \frac{\mathcal F_A(k)}{\mathcal F_p(k)}
\;.
\end{equation}
Here, we denote the proton and nuclear dipole amplitude by $\mathcal F_A(k) $ and $\mathcal F_p(k) $, respectively. Note that by giving the photon and jet kinematics, the small-$x$ scale in the target $x_A$ is determined.
In Fig.~\ref{fig:R_Fkx}, we present the ratio of the Au (Pb) and proton dipole amplitudes at the corresponding $x_A$ used in Fig.~\ref{fig:RHIC_jet_TEEC} [Fig.~\ref{fig:LHC_jet_TEEC}]. 
If, in addition, one fixes the photon and jet $p_T$, the value of $\tau$ is also determined, as the momentum transfer in the dipole amplitude depends solely on $\tau$. 
In both panels, the $\tau=0.1$ and $0.9$ are indicated by red vertical dashed lines, and the intermediate $\tau$s are in between. 

 \begin{figure*}[htp!]
  \centering 
  \subfigure[
    RHIC
    \label{fig:RHIC_jet_ATEEC}
  ]{ \includegraphics[width=0.45\textwidth]{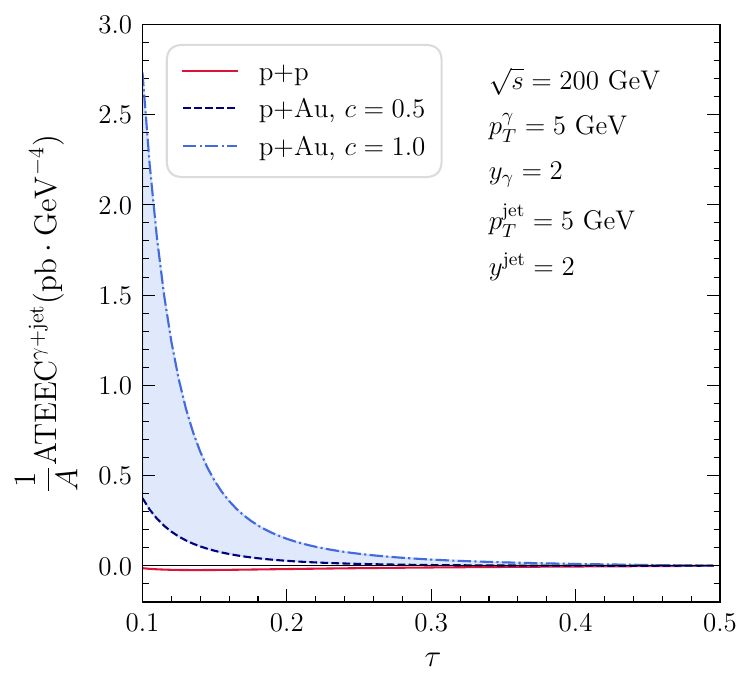} }\qquad
  \subfigure[
    LHC
    \label{fig:LHC_jet_ATEEC}
  ]{ \includegraphics[width=0.45\textwidth]{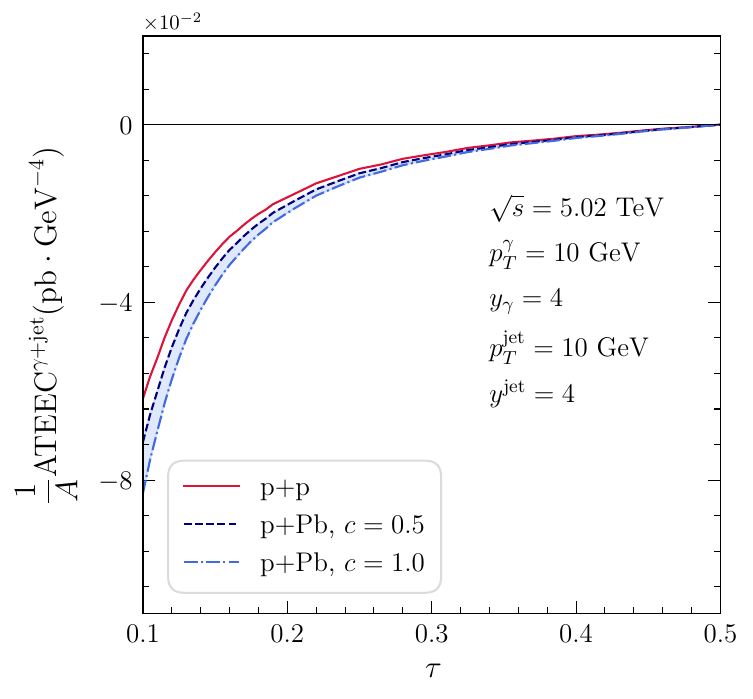} }
  \caption{The photon--jet ATEEC, as defined in Eqs.~\eqref{eq:TEEC_jet} and~\eqref{eq:ATEEC}, as a function of $\tau$.}
  \label{fig:ATEEC_photon_jet}
\end{figure*}

In the left panel, the ratio is below $1$ at very small $k$, increasing and surpassing $1$ at larger $k$ before peaking around $k=1\sim 3 \GeV$ and decreasing towards a nearly constant value close to $1$.
For the MV model, it has been shown that due to the particular $A$ dependence of the dipole--target cross section, 
there is an enhancement in the nuclear modification at large $k$ \cite{Gelis:2002nn, Accardi:2002ik, Kharzeev:2003wz}, a phenomenon known as the Cronin effect \cite{Antreasyan:1978cw}. The location of the peak of the enhancement (the ``Cronin peak'') is associated with the saturation scale. 
In our calculation, the initial saturation scale for the nuclear target (Au) is given by $Q_{s0,A}^2=cA^{1/3}Q_{s0}^2$ according to Eq.~\eqref{eq:Qs_A}. A larger value of $c$ corresponds to scattering at a higher saturation scale, leading to a more pronounced Cronin peak and a shift to a higher peak position.
Specifically, the ratio of the two peak positions expected by the saturation scale is $Q_{s0}(c=1.0)/Q_{s0}(c=0.5)=\sqrt{2}$, which agrees with our findings shown in the left panel of Fig.~\ref{fig:R_Fkx}.
As indicated by the $\tau$ lines in the figure, we are probing the $k$ region after the Cronin peak; therefore, we see an enhanced but decreasing $R_{pA}$ in Fig.~\ref{fig:RHIC_jet_TEEC}. Probing the peak region requires a smaller photon and/or jet $p_T$.
In contrast, the ratio in the right panel is at a much smaller $x_A$, and is below 1 in the entire $k=0\sim \SI{30}{GeV}$ region being plotted, 
and there is no clear Cronin peak anymore.
This flattening of the Cronin peak when going to higher energies is expected~\cite{Kharzeev:2003wz},
and instead of the Cronin peak the behavior of the nuclear modification factor is now similar to the region to the left of the Cronin peak in the left panel: higher saturation scale (the $c=1.0$ case) leads to more suppression.
Interestingly, the curves in the right panel lie below 1 even at high values of $k$, indicating a sizable suppression in the nuclear modification.

Next, in Fig.~\ref{fig:ATEEC_photon_jet} we focus on the $\gamma+\jet$ ATEECs as defined in Eq.~\eqref{eq:ATEEC}. 
At RHIC energy, as shown in Fig.~\ref{fig:RHIC_jet_ATEEC}, the $\gamma+\jet$ TEEC for p+p happens to be relatively symmetric, resulting in an ATEEC curve that stays close to zero.
In contrast, the TEEC for p+A is noticeably larger near the back-to-back region than that in the large $\tau$ region, leading to a positive and decreasing ATEEC curve.
On the other hand, at LHC energy, the nuclear modification factor remains within a relatively flat band, as shown in Fig.~\ref{fig:LHC_jet_TEEC} and the right panel of Fig.~\ref{fig:R_Fkx}. 
Consequently, the ATEECs for p+Pb collisions, as shown in Fig.~\ref{fig:LHC_jet_ATEEC}, exhibit little qualitative difference from that for p+p, which is increasing and concave down in the plotted region. 
However, the asymmetry in the band itself caused by the p+Pb asymmetry is greater in magnitude than for p+p even though the corresponding TEEC is smaller in magnitude, as shown in Fig.~\ref{fig:LHC_jet_TEEC}.

\begin{figure*}[htp!]
  \centering
  \subfigure[ $\gamma+h$ TEEC \label{fig:TEEC_hadron}]{ 
\includegraphics[width=0.4377\textwidth]{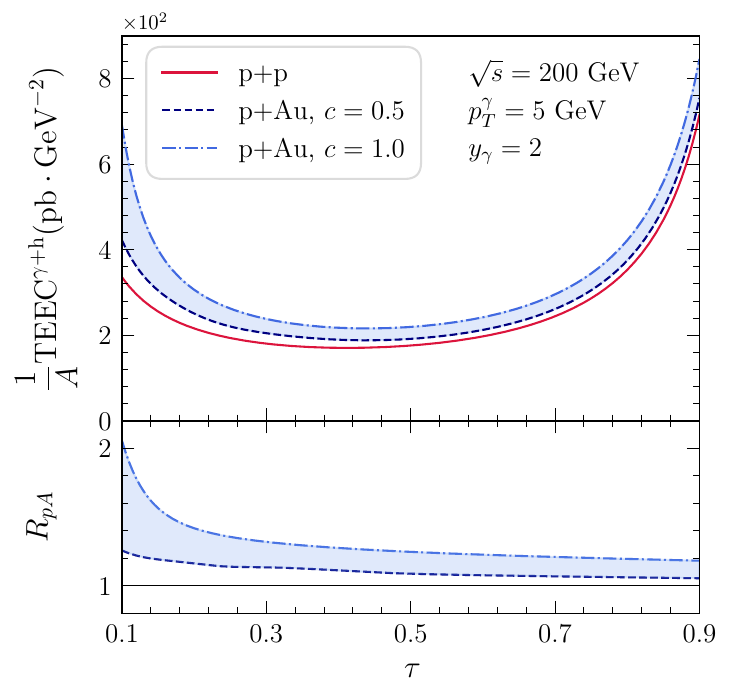}
\includegraphics[width=0.45\textwidth]
{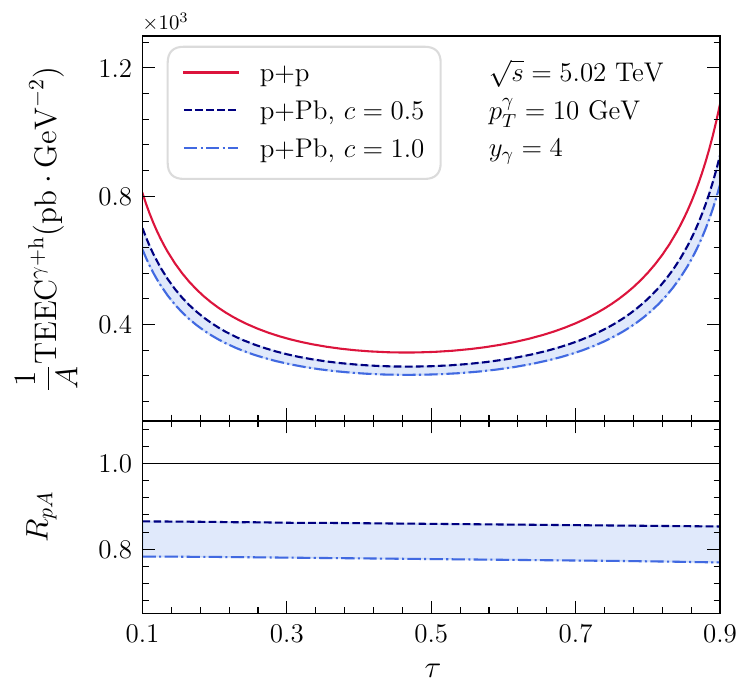}
}
  \subfigure[ $\gamma+h$ TEEC asymmetry \label{fig:ATEEC_hadron}]{ 
  \includegraphics[width=0.45\textwidth]{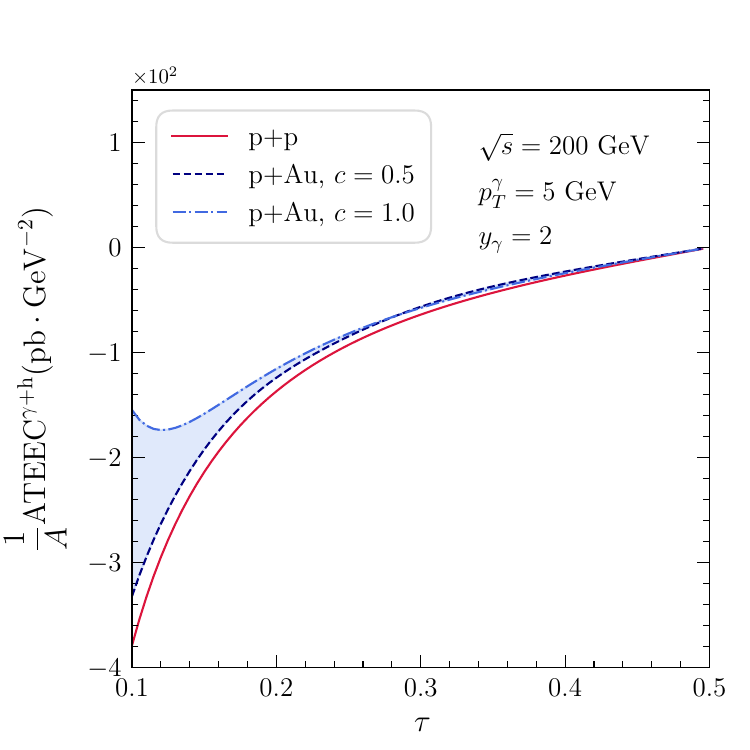}
\includegraphics[width=0.45\textwidth]
{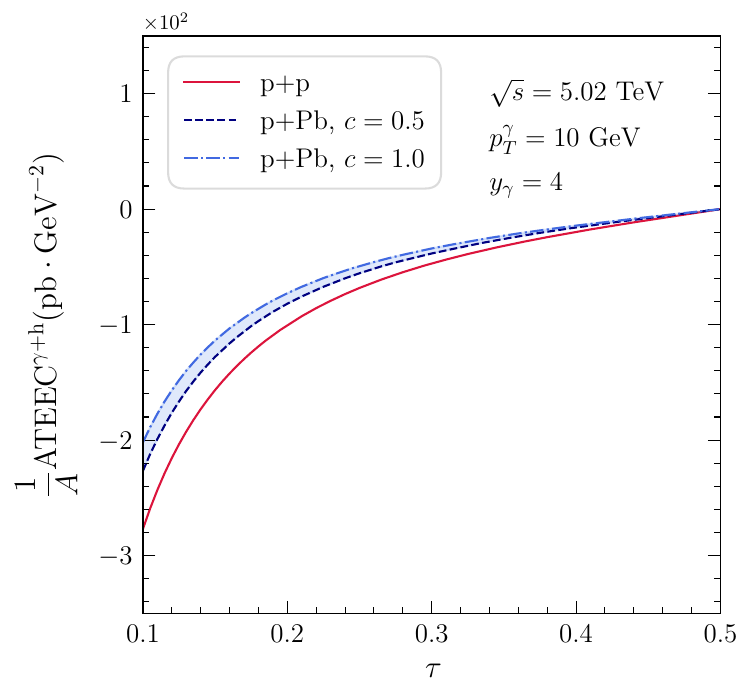}
}
\caption{The photon--hadron TEEC and its asymmetry ATEEC, as defined in Eqs.~\eqref{eq:TEEC_hadron} and~\eqref{eq:ATEEC}, shown as functions of $\tau$.
}
  \label{fig:photon_hadron}
\end{figure*}

\subsection{Photon + hadron}
\label{sec:numerics_hadron}

We calculate the $\gamma+h$ TEEC, as defined in Eq.~\eqref{eq:TEEC_hadron}, together with the associated nuclear modification ratio $R_{pA}$ in the selected RHIC and LHC kinematic regions. The results are presented in Fig.~\ref{fig:TEEC_hadron}.
As in the $\gamma+\jet$ case, the TEECs are larger at the smaller and larger ends of $\tau$, due to the overall factor of $1/\sqrt{\tau(1-\tau)}$ in Eq.~\eqref{eq:TEEC_hadron}. 
The $R_{pA}$ exhibits a similar behavior compared to that of the $\gamma+\jet$ TEEC (as in Fig.~\ref{fig:TEEC_photon_jet}) with the same jet kinematics; it decreases asymptotically to a value slightly above 1 over that $\tau$ range. 
A difference is that the $\gamma+h$ band is slightly wider because it arises from integrating over a range of $x_A$ values. The study in Ref.~\cite{Ganguli:2025aqa} also reports nuclear suppression especially in the small $\tau$ region, albeit at a larger center-of-mass energy. Moreover, the p+p graph is now clearly asymmetric, rising much higher in the forward region ($\tau\to 1$) than in the back-to-back region ($\tau\to 0$).

Figure~\ref{fig:ATEEC_hadron} shows the $\gamma+h$ ATEECs as defined in Eq.~\eqref{eq:ATEEC}. The aforementioned asymmetries in the p+p TEEC curves are reflected in the corresponding ATEEC curves. All three ATEEC curves in the left panel, i.e., at the RHIC energy, remain negative, indicating a larger correlator in the forward region compared to that near the back-to-back region, exhibiting a different behavior compared to that in the $\gamma+\jet$ case, as shown in Fig.~\ref{fig:RHIC_jet_ATEEC}. The three ATEEC curves in the right panel, corresponding to the LHC energy, are also negative---similar to the $\gamma+\text{jet}$ case shown in Fig.~\ref{fig:LHC_jet_ATEEC}---but exhibit a stronger asymmetry in the p+p case compared to p+A.

\section{Conclusions}
\label{sec:conclusions}
We have studied the transverse energy--energy correlator (TEEC) and its derived observables---the nuclear modification factor $R_{pA}$ and the asymmetry ATEEC---in both photon--hadron and photon--jet production in p+p and p+A collisions at small~$x$. We have computed these observables at the RHIC and LHC energies using a framework based on the dipole picture, where nonlinear gluon dynamics in the small-$x$ regime are encoded in the dipole amplitude evolved via the rcBK equation. 

One of the main advantages of the photon--hadron TEEC over standard azimuthal correlations is that the dependence on the fragmentation function cancels due to the momentum-sum rule, thereby reducing theoretical uncertainties associated with hadronization. The TEEC is directly sensitive to the structure of the target in the high-energy limit. We find that for smaller $p_T^\gamma$ and smaller values of $\tau$, the results are more sensitive to variations in the initial condition of the dipole amplitude, as reflected in the uncertainty band for $c \in [0.5, 1.0]$. This is expected, since these regions correspond to lower momentum transfers $\sim 2\sqrt{\tau}\, p_T^\gamma$, which are more affected by gluon saturation effects. Extending the calculation to even smaller values of $\tau$ would further enhance sensitivity to saturation physics but would require the resummation of Sudakov logarithms, which we leave for future work.

The results demonstrate that TEEC and ATEEC are promising observables for probing gluon saturation in high-energy nuclear collisions. The formalism developed here can also be extended to study TEECs in dihadron and dijet production in both p+p and p+A collisions. We look forward to future experimental measurements of TEEC and ATEEC observables, which can further constrain the small-$x$ gluon structure and deepen our understanding of QCD dynamics in the saturation regime.

\section*{Acknowledgments}
We thank Tuomas Lappi, Heikki M\"antysaari, Carlos A. Salgado, Xuan-bo Tong, Bowen Xiao, and Yiyu Zhou, for valuable discussions. 
R.K., Z.K., and J.P. are supported by the National Science Foundation under grant No. PHY-1945471.
This work is also supported by the U.S. Department of Energy, Office of Science, Office of Nuclear Physics, within the framework of the Saturated Glue (SURGE) Topical Theory Collaboration.
M.L. is supported by Xunta de Galicia (CIGUS accreditation), European Union ERDF, the Spanish Research State Agency under project PID2020-119632GB-I00, and European Research Council under project ERC-2018-ADG-835105 YoctoLHC. 

\bibliographystyle{apsrev4-1}

\bibliography{main.bib}

\end{document}